\documentclass[10pt,a4paper]{amsart}
\usepackage{latexsym,amssymb,amsmath,amsthm,amsfonts,enumerate,verbatim,xspace,
exscale}

\usepackage{color,amsmath,amsbsy,amsfonts,amssymb}
 
\usepackage{hyperref}
\definecolor{darkblue}{rgb}{0,0,.5}
\hypersetup{pdftex=true, colorlinks=true, breaklinks=true, linkcolor=darkblue, menucolor=darkblue, pagecolor=darkblue, urlcolor=darkblue,citecolor=darkblue}

\theoremstyle{plain}
\newtheorem{theorem}{Theorem}[section]
\newtheorem{lemma}[theorem]{Lemma}
\newtheorem{proposition}[theorem]{Proposition}
\newtheorem{corollary}[theorem]{Corollary}

\theoremstyle{definition}
\newtheorem{definition}[theorem]{Definition}

\def\D{\mathcal{D}}
\def\C{\mathcal{C}}
\def\R{\mathbb{R}}
\def\I{\mathbb{I}}

\def\M{\mathcal{M}}
\def\N{\mathcal{N}}
\def\F{\mathcal{F}}

\def\A{\mathcal{A}}
\def\Ham{\mathcal{H}}

\newcommand{\caps}[1]{\textup{\textsc{#1}}}

\providecommand{\bysame}{\makebox[3em]{\hrulefill}\thinspace}

\newcommand{\up}{\upshape}

\newcommand{\longto}{\longrightarrow}
\newcommand{\hookto}{\hookrightarrow}
\newcommand{\toto}{\twoheadrightarrow}

\def\vv<#1>{\langle#1\rangle}

\newcommand{\pr}{\mbox{$\text{\up{pr}}$}}

\providecommand{\det}{\mbox{$\text{\up{det}}\,$}}

\newcommand{\dd}[2]{\mbox{$\frac{\partial #2}{\partial #1}$}}

\newcommand{\om}{\omega}
\newcommand{\Om}{\Omega}

\newcommand{\lam}{\lambda}

\newcommand{\wt}[1]{\mbox{$\widetilde{#1}$}}

\newcommand{\by}[2]{\mbox{$\frac{#1}{#2}$}}
\newcommand{\cinf}{\mbox{$C^{\infty}$}}
\providecommand{\set}[1]{\mbox{$\{#1\}$}}

\newcommand{\summ}{\mbox{$\sum$}}

\newcommand{\X}{\mathfrak{X}}

\newcommand{\curv}{\mbox{$\textup{Curv}$}}

\newcommand{\hor}{\mbox{$\textup{Hor}$}}

\newcommand{\momap}{momentum map\xspace}

\newcommand{\gu}{\mathfrak{g}}
\newcommand{\ho}{\mathfrak{h}}

\newcommand{\ko}{\mathfrak{k}}

\newcommand{\Ad}{\mbox{$\text{\upshape{Ad}}$}}

\newcommand{\SO}{\mbox{$\textup{SO}$}}
\newcommand{\so}{\mbox{$\mathfrak{so}$}}

\newcommand{\hlphi}{\mbox{$\textup{hl}^{\mathcal{A}}$}}

\newcommand{\Hamc}{\mbox{$\mathcal{H}_{\textup{c}}$}}
\newcommand{\Omnh}{\mbox{$\Om_{\textup{nh}}$}}

\title[Stochastic Chaplygin systems]{Stochastic Chaplygin systems}

\author{Simon Hochgerner}
\address{Section de Mathematiques,
Station 8,
EPFL, CH-1015 Lausanne}
\email{simon.hochgerner@epfl.ch} 
\date{February 8, 2010}

\begin{document}

\begin{abstract}
We mimic the stochastic Hamiltonian reduction of Lazaro-Cami and
Ortega \cite{LO08,LO08a}
for the case of certain non-holonomic systems with symmetries.

Using the non-holonomic connection it 
is shown that the drift of the stochastically perturbed 
$n$-dimensional Chaplygin
ball is a certain gradient of the density of the preserved measure 
of the deterministic system.
\end{abstract}

\maketitle
\tableofcontents

\section{Introduction} 

Imagine a ball sitting on a rough horizontal table. Because of the
roughness of the table this ball -the
Chaplygin ball- cannot slip,
but it can turn about the vertical axis without
violating the constraints. Its geometric and gravitational center
coincide but there may be an inhomogeneous mass distribution. 
Suppose
now that the ball is subjected to a Brownian noise such that there is
a random jiggling in all of its angular and translational degrees of
freedom. 
This problem
is similar to the stochastic rigid body considered in \cite{LO08a} but
upon imposing the no-slip constraints some differences can be
expected. 
One may
wonder if the stochastic Chaplygin 
ball will acquire a drift that makes it roll on the
table or spin about its vertical axis or both?  
The answer to this question is given by Theorem~\ref{thm:diff}: The
drift follows a Fick's law in the following sense. 
The configuration space of the $n$-dimensional ball is
$Q = \SO(n)\times\R^{n-1}$ and there are $n-1$ constraints corresponding
to the directions in the table $\R^{n-1}$. 
By a symmetry
reduction argument (compression) 
one can eliminate the $\R^{n-1}$-factor and
the \emph{deterministic} motion of the ball can be described by the geodesic
equations of the so-called non-holonomic connection
$\nabla^{\textup{nh}}$ on $\SO(n)$. With respect to this connection
one can now show that the process on $\SO(n)$ describing
the balls \emph{stochastic} motion is a non-holonomic
diffusion separating 
into a drift- and a martingale-term.
See Definition~\ref{def:drift}. 
Theorem~\ref{thm:diff} says that this $\nabla^{\textup{nh}}$-drift equals 
\[
-\by{1}{2}\textup{grad}^{\mu_0}\log\N
\]
where $\N$ is the preserved density \eqref{e:N} of the
deterministic ball and the gradient is computed with respect to the
kinetic energy metric \eqref{E:mu_0} of the ball.

In particular, when the ball is homogeneous there is no drift. 
Moreover, it is shown in Corollary~\ref{cor:hom-case} that the
homogeneous ball's
stochastic process
factorizes to a Brownian motion on `the ultimate reduced configuration
space' $S^{n-1} = (Q/\R^{n-1})/\SO(n-1)$. This is in analogy to the
corresponding deterministic case where the motion is Hamiltonian at the ultimate
reduced level. (See \cite{HG09}.) 


In $3$ dimensions the drift 
$-\by{1}{2}\textup{grad}^{\mu_0}\log\N$
does not have an angular velocity
component about the vertical axis of the ball in the space frame. 
For dimensions $n>3$ it turns out that 
this property is related to the Hamiltonization of the
deterministic system: If the inertia matrix describing the
balls mass distribution satisfies the Hamiltonization condition
\ref{e:Ham-cond} then the drift does not have an angular velocity
component about the vertical axis in the space frame.
On the other hand, the drifts angular momentum about the vertical axis
is always $0$, regardless of the dimension or the mass distribution.

Section~\ref{sec:2} starts by collecting some definitions and
facts from stochastic differential geometry as presented in
\cite{IW89,E89}. Then we rehearse the basics of the stochastic
Hamiltonian mechanics and their symmetries as introduced in
\cite{LO08,LO08a}. In Section~\ref{sec:stoch-Chap} these ideas are
transferred to describe stochastic $G$-Chaplygin systems. It is
noticed that the reduction of symmetries, termed \emph{compression} in
this context, works naturally.

In Section~\ref{sec:3} this construction is applied to the
$n$-dimensional Chaplygin ball. First some facts about the
deterministic system, such as the preserved measure and
Hamiltonization, are recalled. Then the Hamiltonian construction of
\cite{LO08} of Brownian motion on the configuration space $Q =
\SO(n)\times\R^{n-1}$ is reviewed. Section~\ref{sec:comp-of-BM}
makes the constraint forces act on the Brownian motion according to
the recipe of Section~\ref{sec:stoch-Chap}.
Thus a constrained stochastic motion is obtained and the above
mentioned Theorem~\ref{thm:diff} is found.
Finally, we note that we can also treat the cases of angular jiggling
only or horizontal jiggling only. The latter corresponds to the
Chaplygin ball
sitting on a table which undergoes a translational Brownian motion. 
The $3$-dimensional version of this case has been considered in
\cite{MS90}.

\section{Stochastic Chaplygin systems and reduction of symmetries}\label{sec:2}



\subsection{Stochastic geometry}

We state some notions from stochastic differential geometry. The
references we used here are \cite{E89,IW89}. See also the appendix of
\cite{LO08}. 

Let $M$ be a manifold,
let $(\Om,\F,\set{\F_t:t\ge0},P)$ be
a filtered probability space, and let $\Gamma: \R_+\times\Om\to M$ be
an adapted
stochastic process. (We consider only continuous processes.)

The process $\Gamma$ is called a \caps{semi-martingale} if
$f\circ\Gamma$ is a semi-martingale in $\R$ for all
$f\in\cinf(M)$. See \cite[Chapter~III]{E89}.

The definition of a martingale in $M$ depends upon a choice of a
connection.
Let $\nabla$ be a 
connection in $TM\to M$. 
Then
the Hessian of $\nabla$ is defined by
\[
\textup{Hess}^{\nabla}(f)(X,Y)
= 
XY(f)-\nabla_XY(f)
\]
for $X,Y\in\X(M)$.
This is bilinear in $X$ and $Y$ but not symmetric, in general, 
since
$\textup{Hess}^{\nabla}(f)(X,Y) 
- \textup{Hess}^{\nabla}(f)(Y,X)
= -\textup{Tor}^{\nabla}(X,Y)(f)$.

\begin{definition}\label{def:mart}
A semi-martingale $\Gamma: \R_+\times\Om\to M$ 
is said to be a
$\nabla$-\caps{martingale}
if, for any $f\in\cinf(M)$,
\[
 f\circ\Gamma_t - f\circ\Gamma_0 
 -
 \int_0^t \textup{Hess}^{\nabla}(f)(d\Gamma_s,d\Gamma_s)ds
\]
is a local martingale in $\R$.
\end{definition}

In \cite[Chapter~IV]{E89} this is stated in terms of
a torsionless connection but it is noted that one can
also allow for connections with torsion, since it is proved
(\cite[(3.14)]{E89}) that 
$\int_0^t \textup{Hess}^{\nabla}(f)(d\Gamma_s,d\Gamma_s)ds$ depends
only on the symmetric part of $\textup{Hess}^{\nabla}$.
The situation is similar to the notion of a $\nabla$-geodesic. This
also depends only on the torsionless part 
$\nabla-\by{1}{2}\textup{Tor}^{\nabla}$
of $\nabla$.

Let $M$ and $N$ be manifolds, let $(\Om,\F,\set{\F_t:t\ge0},P)$ be
a filtered probability space, and let $X: \R_+\times\Om\to N$ be a
semi-martingale. A \caps{Stratonovich operator} $\mathcal{S}$ from $TN$ to $TM$ is a family
of linear linear maps 
\[
 \mathcal{S}_{(x,y)}: T_xN\longto T_yM
\]
which depends smoothly on $x\in N$ and $y\in M$.
In other words, $\mathcal{S}$ is a section of $T^*N\otimes TM\to
N\times M$. A Stratonovich
differential equation for a semi-martingale $\Gamma: \R_+\times\Om\to
M$ is written as 
\[
 \delta\Gamma = \mathcal{S}(\Gamma,X)\delta X.
\]
See \cite[Chapter~VII]{E89} for the precise meaning of this equation 
as well as existence and
uniqueness (up to explosion time) of solutions.  

Assume now that $N=\R\times\R^n$ and that $X:
\R_+\times\Om\to\R\times\R^n$, $(t,\om)\mapsto(t,W_t(\om))$ where $W$
denotes $n$-dimensional Brownian motion. Let $X_0,X_1,\dots,X_n$ be
vectorfields on $M$ and define the Stratonovich operator
\[
 \mathcal{S}(x,y): \R^{n+1}\longto T_yM,\;
 (t,w^1,\dots,w^n)\longmapsto tX_0(y) + \summ w^iX_i(y).
\]
Let $f\in\cinf(M)$.
Then
$f\circ\Gamma: \R_+\times\Om\to\R$ satisfies
\[
 \delta(f\circ\Gamma)
 = df(\Gamma).\mathcal{S}(X,\Gamma)\delta X
 = df(\Gamma).X_0(\Gamma)\delta t
   + \summ df(\Gamma).X_i(\Gamma)\delta W^i.
\]
Hence $\Gamma$ defines a diffusion in $M$ by 
\cite[Capter~V, Thm.~1.2]{IW89}. 
The generator of this diffusion is the second order differential
operator $A$ given by
\[ 
 Af = X_0f + \by{1}{2}\summ X_iX_i f
\]
where $f\in\cinf(M)$.

\begin{definition}\label{def:BM}
If $M$ is equipped with a Riemannian metric $\mu$ and $\Delta$ is the
associated Laplacian then the diffusion is called 
\caps{Brownian motion} in $(M,\mu)$ if 
\[
 A = \by{1}{2}\Delta.
\]
\end{definition}

This definition agrees with the one given in \cite[(5.16)]{E89} or
\cite[(8.5.18)]{Oks07} but differs slightly from the one in
\cite[Chapter~V,~Def.~4.2]{IW89} where it is required that 
$n=\dim M$.

\subsection{Stochastic Hamiltonian systems}

This section presents some of the concepts elaborated in
\cite{LO08,LO08a}. 

Let again $N=\R^n$ and consider a Poisson manifold $(M,\{.,.\})$
together with an $\R^n$-valued Hamiltonian function
$h=(h_1,\dots,h_n): M\to\R^n$. 
Let $X: \R_+\times\Om\to\R^n$ be a semi-martingale. 
The associated stochastic Hamiltonian
system is given by the Stratonovich equation 
$\delta\Gamma = \mathcal{S}(X,\Gamma)\delta X$
where
$\mathcal{S}$ is defined in terms of the Hamiltonian structure, that
is,
\[
 \mathcal{S}(x,y): 
 \R^n\longto T_yM,\;
 (x^1,\dots,x^n)
 \longmapsto
 \summ X_{h_i}(y)x^i
\]
where the Hamiltonian vectorfield $X_{h_i}$ is the vectorfield
corresponding to the derivation $\{f,.\}$. 
When $(M,\om)$ is a symplectic manifold then one uses the Hamiltonian
fields defined by $i(X_{h_i})\om = dh_i$. 

These systems allow for a 
symmetry reduction analogous to classical mechanics. We state the
symplectic version of this
theorem of \cite[Section~6]{LO08a}:

\begin{theorem}[Stochastic Hamiltonian reduction]\label{thm:stoch-ham-red}
Let $(M,\om)$ be a symplectic manifold with Hamiltonian $h: M\to\R^n$ and
stochastic component $X$ as above. Assume that $(M,\om,h)$ are
invariant under the free and proper action of a Lie group $G$ such
that a coadjoint equivariant \momap $J: M\to\gu^*$ exists. Fix a level
$\lam\in\gu^*$. 

Then $J^{-1}(\lam)$ is invariant under the flow of
$\mathcal{S}$. Moreover, $\mathcal{S}$ induces a Stratonovich operator
$\mathcal{S}_{\lam}$ from $T\R^n$ to $T(J^{-1}(\lam)/G_{\lam})$ and solutions of
$\mathcal{S}$ with initial condition in $J^{-1}(\lam)$ project to
solutions of $\mathcal{S}_{\lam}$. The induced operator is given by
\[
 \mathcal{S}_{\lam}(x,y_0): 
 (x^1,\dots,x^n)\longmapsto \summ X_{h^{\lam}_i}(y_0)x^i 
\]
where $x\in\R^n$, $y_0\in J^{-1}(\lam)/G_{\lam}$ and $X_{h^{\lam}_i}$
is the Hamiltonian vectorfield with
respect to the reduced symplectic form
on $J^{-1}(\lam)/G_{\lam}$ of the induced function
$h^{\lam}_i$. 
\end{theorem}

\subsection{Stochastic $G$-Chaplygin systems}\label{sec:stoch-Chap}

A (deterministic) $G$-Chaplygin system consists of a Riemannian configuration manifold
$(Q,\mu)$, a Lie group $G$ acting freely and properly by isometries on
$(Q,\mu)$, and a horizontal space $\D$ of the principal bundle $\pi: Q\toto
Q/G$.
Hence $\D$ is the kernel of a connection form $\A: TQ\to\gu$.
 The Lagrangian of the system is the kinetic energy, i.e.,
$L(q,v) = \by{1}{2}\mu_q(v,v)$. In general, $\D$ is \emph{not}
$\mu$-orthogonal to the vertical space $\ker T\pi$. We will henceforth
identify $TQ=T^*Q$ via $\mu$. 
See also Section~\ref{sec:Chap}.

Let $J_G: TQ\to\gu^*$ denote the standard equivariant \momap
associated to the lifted $G$-action on $TQ$.
Given a $G$-invariant function $h: TQ\to\R$ one may use Noether's 
theorem to conclude that the Hamiltonian vectorfield $X_h$ is tangent
to $J_G^{-1}(0)$ and, moreover, is projectable for $J_G^{-1}(0)\toto
J_G^{-1}(0)/G = T(Q/G)$. 
This is why results such as Theorem~\ref{thm:stoch-ham-red} work. 
Since $\D$ is a perturbed version of $J_G^{-1}(0)$ we
look for a substitute construction.

Consider the horizontal space associated to the pulled back connection
$\tau^*\A = \A\circ T\tau: T(TQ)\to TQ \to\gu$ of the tangent lifted $G$-action
on $TQ$,
\[
 \F := \ker\tau^*\A \subset TTQ.
\]
By assumption $\D$ is also $G$-invariant. Thus we can consider the
restricted $G$-action on $\D$ and the associated connection
$\iota^*\tau^*\A: T\D\to\gu$ where $\iota: \D\hookto TQ$ is the
inclusion. Define
\[
 \C := \ker\iota^*\tau^*\A \subset T\D
\]
to be the horizontal space of the principal bundle $\D\toto\D/G = T(Q/G)$

According to \cite{BS93} we can decompose
$\F$ along $\D$ as 
\[
 \F|\D = \C\oplus(\F|\D)^{\Om}
\]
where $(\F|\D)^{\Om}$ is the $\Om$-orthogonal of $\F|\D$ in $TTQ|\D$.
In particular, the fiber-wise restriction of $\iota^*\Om$ to
$\C\times\C$ is non-degenerate. 
For $z\in\D$ define the projection
\begin{equation}\label{e:P}
 P_z: T_z(TQ)\longto\F_z\longto\C_z
\end{equation}
where we first project along the vertical space of the $G$-action on
$TQ$ and then along $\F_z^{\Om}$.  
Moreover, for a $k$-form $\phi$ on $TQ$ we denote the fiber-wise
restriction of $\iota^*\phi$ to $\Pi^k\C$ by
$\phi^{\mathcal{C}}$.
For a function $h: TQ\to\R$ we may thus define the vectorfield
$X^{\mathcal{C}}_h$ on $\D$ with values in $\C$ by the formula
\begin{equation}\label{e:XC}
 P_z X_h(z) 
 =
 (\Om^{\mathcal{C}})_z^{-1}(dh)^{\mathcal{C}}_z
 =:
 X_h^{\mathcal{C}}(z)
\end{equation}
where $z\in\D$.
The link to to the non-holonomic system introduced above is
the following: Via the Legendre transform the dynamics of the non-holonomic system
$(Q,\D,L)$ can be equivalently described by the triple
$(TQ,\Om^C,\mathcal{H}=\by{1}{2}\mu(p,p))$ together with equation
\eqref{e:XC}.
Thus the Lagrange multipliers have been encoded in the two-form
$\Om^{\mathcal{C}}$
or, equivalently, in the projection $P: T(TQ)|\D\to\C$.
The idea is that a non-holonomic system is a Hamiltonian system acted
upon by constraint forces. The effect of the forces is described by the projector
$P: T(TQ)|\D\to\C$. 

In addition to this structure
consider now a semi-martingale $X: \R_+\times\Om\to\R^n$ and a Hamiltonian
function $h=(h_i)_i: TQ\to\R^n$ as above. The associated stochastic non-holonomic
system is given by the Stratonovich equation $\delta\Gamma =
\mathcal{S}^{\mathcal{C}}(X,\Gamma)\delta X$ where 
the Stratonovich operator 
$\mathcal{S}^{\mathcal{C}}: \R^n\times\D\to\textup{Hom}(T\R^N,\C)$  
is given by
\begin{equation}\label{e:stoch-Str}
 \mathcal{S}^{\mathcal{C}}(n,z): 
 \R^n\longto \C_z,\;
 (x^1,\dots,x^n)\longmapsto
   \summ P_z X_{h_i}(z)x^i = \summ X^{\mathcal{C}}_{h_i}(z)x^i.
\end{equation}
Thus the non-holonomic Stratonovich operator arises, by applying the
constraint forces, as a projection of the
Hamiltonian Stratonovich operator into $\C$. 
When the $h^i$ are $G$-invariant we refer to the collection
$(Q,\D,h=(h_i)_i,X)$ as a stochastic $G$-Chaplygin system.

\begin{proposition}[Compression of stochastic $G$-Chaplygin systems]
Let $(Q,\D,h=(h_i)_i,X)$ be a stochastic $G$-Chaplygin
system with $\Om^{\mathcal{C}}$ as above. 
Then the Stratonovich operator \eqref{e:stoch-Str}
compresses to a Stratonovich operator $\mathcal{S}^{\textup{nh}}$ from
$T\R^n$ to 
$\D/G = T(Q/G)$ which is given by 
\[
  \mathcal{S}^{\textup{nh}}(n,z_0): 
 \R^n\longto T_{z_0}(T(Q/G)),\;
 (x^1,\dots,x^n)\longmapsto
   \summ X^{\textup{nh}}_{h^0_i}(z)x^i 
\]
where $z_0\in T(Q/G)$ and $h_i^0: T(Q/G)\to\R$ is the function induced
on the quotient from the invariant function $\iota^*h_i$. Moreover,
solutions of \eqref{e:stoch-Str} project to solutions of $\mathcal{S}^{\textup{nh}}$.
\end{proposition}

\begin{proof}
Everything is entirely analogous to the proof of
\cite[Theorem~6.7]{LO08a} with the only difference that now one uses
Proposition~\ref{prop:comp} instead of the usual symplectic reduction
theorem.
\end{proof}

We think of 
$\delta\Gamma =
\mathcal{S}^{\textup{nh}}(X,\Gamma)\delta X$
as the equations of motion of the 
 system
$(Q,\D,h,X)$.

\begin{proposition}[Ito representation]\label{prop:nh-Ito}
Let $f\in\cinf(T(Q/G))$. Then the Ito representation of 
the equation
$\delta(f\circ\Gamma) =
\mathcal{S}^{\textup{nh}}(X,f\circ\Gamma)\delta X$ is
\[
 d(f\circ\Gamma)
 =
 \sum_i X^{\textup{nh}}_{h_i^0}f(\Gamma)dX^i 
 +
 \by{1}{2}\sum_{i,j}X^{\textup{nh}}_{h_i^0}X^{\textup{nh}}_{h_j^0}f(\Gamma)[dX^j,dX^i]
\]
where $X^i=\pr^i\circ X$.
\end{proposition}

\begin{proof}
This follows exactly as in the proof \cite[Proposition~2.3]{LO08}. It
is only necessary to notice that this proof does not depend on whether
or not the non-holonomic bracket 
$X^{\textup{nh}}_gf =
\{g,f\}^{\textup{nh}} = -\{f,g\}^{\textup{nh}}$ 
satisfies the Jacobi identity. 
The only property of the Poisson bracket which is used in
\cite[Proposition~2.3]{LO08} is the Leibniz rule and this 
feature is evidently shared by the non-holonomic bracket. 
\end{proof}

\section{The stochastic Chaplygin ball}\label{sec:3}

\subsection{The deterministic system}
For background on the Chaplygin ball we refer to
\cite{C02,D04,EKMR04,FK95,HG09,Jov09}. 
The configuration space of Chaplygin's $n$-dimensional rolling ball is
$Q=K\times V$ where $K=\SO(n)$ and $V=\R^{n-1}$. The no-slip
constraints are given by the distribution $\D =
(\A+\pr_2)^{-1}(0)\subset TQ$ where 
\[
 \A = \summ \eta^a\otimes e_a
\]
where $e_1,\dots,e_{n-1}$ is the standard basis on $V$ 
and we stick to the following conventions:\footnote{In Section~\ref{sec:stoch-Chap}
it was the connection form $\A+\pr_2$ which was called $\A$.}
$TK=K\times\ko$ is trivialized via left-multiplication and $\ko$ is
equipped with the $\Ad$-invariant inner product
$\vv<.,.>$.\footnote{The space $\X_L(K)$ of left invariant
  vectorfields and the Lie algebra $\ko$ will also be identified
  without further notice.} 
Let $H=\SO(n-1)\subset K$ be the
stabilizer in $K$ of the $n$-th standard vector $e_n$  such that $H$
acts on $V$ in the natural way. We decompose $\ko =
\ho\oplus\ho^{\bot}$ 
with respect to the $\Ad$-invariant inner product $\vv<.,.>$ on
$\ko$. With respect to this inner product we introduce an orthonormal
system 
\[
 Y_{\alpha},\, \alpha = 1,\dots,k = \dim\ho
 \textup{ and }
 Z_a,\,  a = 1,\dots,n-1
\]
on $\ko$ such that $Y_{\alpha}\in\ho$ and $Z_a\in\ho^{\bot}$. 
Associated to this basis we define the right invariant vector fields
$\xi_{\alpha}$ and $\zeta_a$. In the left trivialization these read
\[
 \xi_{\alpha}(s) = \Ad(s^{-1})Y_{\alpha}
 \textup{ and }
 \zeta_a(s) = \Ad(s^{-1})Z_a.
\]
Dually we introduce the corresponding right invariant coframe  
\[
 \rho^{\alpha} = \vv<\xi_{\alpha}, . >
 \textup{ and }
 \eta^a = \vv<\zeta_a, . >.
\] 
The Lagrangian is the function
\[
 L : TQ = K\times \ko\times TV\longto\R,\,
    (s,u,x,x')\longmapsto\by{1}{2}\vv<\I u,u> + \by{1}{2}\vv<x',x'>.
\]
where $\I$ is the
inertia matrix in body coordinates.
The rolling ball with the no-slip constraint is the non-holonomic
system described by the data $(Q,\D,L)$ where the equations of motion
follow from the Lagrange-d'Alembert principle. 
However, we will not have much use for the Lagrange function below
since we will only perturb the resting ball. 
Note also that we overload the symbol $\vv<.,.>$ by using it for the Euclidean
inner product on $V$ as well as for the $\Ad$-invariant structure on
$\ko$.

From a structural point of view the decisive feature of the Chaplygin
ball is that its constraints are given by a
connection $\pr_2+\A: TQ\to V$ on the (trivial) principal bundle
$V\hookto K\times V\toto K$ where $V$ acts on itself by addition. Thus
$\D$ is the horizontal space of this connection. However, $\D$ is not
$\mu$-orthogonal to the vertical space of the bundle.
The
fact that the system is non-holonomic is reflected in the
non-flatness, 
$\curv^{\mathcal{A}} = d\A \neq 0$.

Compression of the Chaplygin ball system yields the almost Hamiltonian
system $(TK,\Omnh,\Hamc)$ where $\Omnh$ is described in
Proposition~\ref{prop:comp}, $TK$ and $T^*K$ are identified via the
induced metric
\begin{equation}\label{E:mu_0}
 \mu_0(u_1,u_2)_s
 =
 \vv<\I u_1,u_2> + \vv<\A_s(u_1),\A_s(u_2)>
 =
 \vv<(\I+\A_s^*\A_s)u_1,u_2>,
\end{equation}
and $\Hamc = \by{1}{2}\mu_0(u,u)_s$. 
The metric $\mu_0$ is the sum of a left invariant and a right
invariant term. Thus it constitutes an $L+R$-system, see \cite{Jov09b}.
Note the useful formula
$\A^*\A(u) = \summ\vv<\zeta_a,u>\zeta_a$. 

The compressed system $(TK,\Omnh,\Hamc)$ is further invariant under the
lift of the left multiplication action of $H$ on $K$. Physically this corresponds
to rotation of the ball about the $e_n$-axis in the space frame. 
This is an \emph{inner symmetry} and gives, 
by the non-holonomic Noether theorem, rise to a
conserved quantity. This quantity is just the standard 
\momap $J_H:
TK\to\ho^*=\ho$, $(s,u)\mapsto\summ\rho^{\alpha}_s(\I u)Y_{\alpha}$ 
of the (co-)tangent lifted $H$-action on $TK$.

The Chaplygin ball shares an important feature with Hamiltonian
systems. Namely, it possesses a preserved measure (\cite{C02,FK95}).
At the compressed level -the $TK$-level-  
the density $\N: K\to\R$ of this measure with respect
to the Liouville volume on $TK$ is 
\begin{equation}\label{e:N}
 \N(s) = (\det\mu_0(s))^{-\frac{1}{2}}.
\end{equation}
This function plays the central role in all questions of
Hamiltonization of the system. Note that $\N$ is $H$-invariant and
thus descends to a function $\N: K/H = S^{n-1}\to\R$.
For further reference we also record that
\begin{equation}\label{e:dN}
 d(\log\N) 
 =
 \summ\vv<[\mu_0^{-1}\zeta_a,\zeta_a],\zeta_b>\eta^b.
\end{equation}

In \cite{HG09} it is proved that $\Omnh$ can be replaced by $\wt{\Om}
= \Om^K - \by{1}{2}\summ\vv<[\zeta_a,\zeta_b],\_>\eta^a\wedge\eta^b$
without altering the equations of motion, that is,
$i(X^{\textup{nh}}_{\mathcal{H}_c})\Omnh 
 = i(X^{\textup{nh}}_{\mathcal{H}_c})\wt{\Om}
 = d\Hamc$.
Now the new system $(TK,\wt{\Om},\Hamc)$ has the same dynamics but
has the advantage of being liable to reduction with respect to the
internal symmetry group $H$: Let $J_H: TK\to\ho^*$ be the
\momap introduced above, $\lam\in\ho^*$ and $\iota: J_H^{-1}(\lam)\hookto TK$ the
inclusion. Then $\iota^*\wt{\Om}$ descends to an almost symplectic two
form $\wt{\Om}_{\lam}$ on $J_H^{-1}(\lam)/H_{\lam}$. 
($\iota^*\Omnh$ is not horizontal for the projection 
onto $J_H^{-1}(\lam)/H_{\lam}$.)
In this way one
can recover the Hamiltonization of the $3$-dimensional ball 
of \cite{BM01,BM05}: It is shown in
\cite[Proposition~4.4]{HG09} that $d(\N\wt{\Om}_{\lam})=0$ if
$n=3$. Thus, for $n=3$, the rescaled vectorfield
$\N^{-1}X^{\textup{nh}}_{\mathcal{H}_c}$ is Hamiltonian with respect
to $(J_H^{-1}(\lam)/H_{\lam}=TS^2,\N\wt{\Om}_{\lam},\Hamc)$. Moreover,
the homogeneous ball, $\I = 1$, is Hamiltonian at the
$J_H^{-1}(\lam)/H_{\lam}$-level for any dimension $n$. It is
interesting to notice that none of these statements hold at the
$TK$-level. 

In \cite[Section~4.A]{H09} the following Hamiltonization condition is
proved for arbitrary dimension $n$:
Let $\lam=0\in\ho^*$. Then $d(\N\wt{\Om}_0) = 0$ if and only if 
\begin{equation}\label{e:Ham-cond}
 (n-2)\vv<\mu_0^{-1}\zeta_d,[\zeta_b,\zeta_c]>
 =
 \summ_a\vv<\mu_0^{-1}\zeta_a,[\zeta_b,\zeta_a]\delta_{c,d}
                             - [\zeta_c,\zeta_a]\delta_{b,d}> 
\end{equation}
which is an algebraic condition on $\I$. For $n=3$ this condition is
trivially satisfied. In the stochastic context this condition appears
in Theorem~\ref{thm:diff} below.

\subsection{Brownian motion on the configuration space}

We follow \cite{LO08} to construct Brownian motion on $Q=K\times
V$. Let $\nabla^{\mu}$ be the Levi-Civita connection of
\[
 \mu((u,x'),(v,y'))
 = 
 \vv<\I u,v> + \vv<x',y'>.
\]
Thus for $u,v\in\ko$ we have 
\[
 \nabla^{\mu}_uv
 =
 (\nabla^{\mathbb{I}}_uv,0) 
 =
 (\by{1}{2}[u,v] + \by{1}{2}\I^{-1}([u,\I v]+[v,\I u]),0)
\]
where $\nabla^{\mathbb{I}}$ is the Levi-Civita connection of
the left-invariant metric defined by $\vv<\I .,.>$ on $K$. 
Note that we identify
$\ko\cong\ko^*$ via $\vv<.,.>$ whence 
$\I: \ko\to\ko^*\cong\ko$. 
Let 
\[
 v_1,\dots,v_m, \, m = \dim\ko = \by{1}{2}n(n-1)
\]
denote a basis which is orthonormal for $\vv<\I .,.>$. Thus
$v_1,\dots,v_m,e_1,\dots,e_{n-1}$ is a left invariant frame on $Q$ 
which is orthonormal with respect to the left invariant metric $\mu$.  
(Remember that we identify $TQ$ and $T^*Q$ via $\mu$.)
Let the functions $H_0,H_i,F_a: TQ\to\R$ be given by
\begin{equation}\label{e:h-f}
 H_0(s,u,x,x') 
 = -\by{1}{2}\vv<\I u, \summ\nabla^{\mathbb{I}}_{v_ i}v_i> ,\;
 H_i(s,u,x,x')
 =
 \vv<\I u,v_i>,
 \textup{ and }
 F_a(s,u,x,x') 
 = \vv<x',e_a>.
\end{equation}
Consider the semi-martingale
\[
 X: \R_+\times\Omega\longto\R\times\R^m\times\R^{n-1},\;
 (t,\om)\longmapsto(t,B^1_t(\om),\dots,B^m_t(\om),W^1_t(\om),\dots,W^{n-1}_t(\om))
\]
where $B^i,W^a$ are $m+n-1$ independent Brownian motions. 
The Stratonovich stochastic differential equation which is
associated to these data is 
\begin{equation}\label{e:Str}
 \delta\Gamma = \mathcal{S}^{\textup{Ham}}(X,\Gamma)\delta X
\end{equation}
where the Stratonovich operator from $T(\R\times\R^m\times\R^{n-1})$
to $TK\times TV$
is defined by
\[
 \mathcal{S}^{\textup{Ham}}(n;s,u,x,x')(t,b^1,\dots,b^m,w^1,\dots,w^{n-1})
 =
 X_{H_0}(s,u,x,x')t
 + \summ X_{H_i}(s,u,x,x')b^i
 + \summ X_{F_a}(s,u,x,x')w^a
\]
with $X_H$ denoting the canonical Hamiltonian vectorfield of a
function $H: TQ\to\R$.
Using the Ito representation of this equation 
\cite{LO08} show that the solutions
$\Gamma$ project via $\tau: TQ\to Q$ onto Brownian motion on $Q$. 

Using again the setting of \cite{LO08} and
Theorem~\ref{thm:stoch-ham-red} it is easy to see the
following. Consider the $V$-action on $Q$ as above. Let 
$J_V^{-1}(0) = \set{(s,u,x,0)}$ be the $0$-level set of the standard
\momap $J_V: TQ\to V^*=V$ of the lifted $V$-action on $TQ$. 
Then the Stratonovich equation \eqref{e:Str}
induces a Stratonovich equation $\delta\Gamma_0 =
\mathcal{S}_0(X,\Gamma_0)\delta X$ on $J_V^{-1}(0)/V = TK$ and
solutions $\Gamma$ with initial condition in $J_V^{-1}(0)$ 
project onto solutions $\Gamma_0$. Moreover,
$\tau_K\circ\Gamma_0$ is a Brownian motion on $K$ where $\tau_K:
TK\to K$. 

When we regard $\D$ as perturbed version of $J_V^{-1}(0)$ we can ask
how much of this observation remains true? This is the content of
Section~\ref{sec:comp-of-BM}.

\subsection{Constrained Brownian motion and compression}\label{sec:comp-of-BM}

We now force the Brownian motion on $Q$ to satisfy the constraints
induced by $\D$. In accordance with Section~\ref{sec:stoch-Chap} we 
do so by applying the constraint forces to the Stratonovich operator
from \eqref{e:Str}. 
Thus we are concerned with the equation 
$\delta\Gamma =
\mathcal{S}^{\textup{nh}}(X,\Gamma)\delta X$ where $X$ and $H_0,H_i,F_a$ are as above and 
\[
 \mathcal{S}^{\mathcal{C}}(n,z) = P_z\mathcal{S}^{\textup{Ham}}(n,z):
  \R\times\R^m\times\R^{n-1}\longto\C_z\subset T_z\D
\]
where $P$ was
defined in \eqref{e:P} and $z=(s,u,x,-\A_s(u))\in\D$.
The functions $H_0,H_i,F_a$ are $V$-invariant and compress to
functions $h_0,h_i,f_a$ given by
\begin{align*}
 h_0(s,u) 
 &= H_0(s,u,x,-\A_s(u)) 
 = -\by{1}{2}\mu((u,-\A_s(u)),\summ\nabla^{\mu}_{v_i}v_i)\\
 &= -\by{1}{2}\vv<\I u,\summ\nabla^{\mathbb{I}}_{v_i}v_i>
 = -\by{1}{2}\mu_0(u,\mu_0^{-1}\I\summ\nabla^{\mathbb{I}}_{v_i}v_i),\\
 h_i(s,u)
 &= \vv<\I u,v_i> 
 = \mu_0(u,\mu_0^{-1}\I v_i),\\
 f_a(s,u)
 &= -\vv<\A_s(u),e_a>
 = -\mu_0(u,\mu_0^{-1}\zeta_a(s)).
\end{align*}
Notice that $h_0$ and $h_i$ are left invariant while the $f_a$ are right
invariant. 
The compressed non-holonomic Stratonovich operator is now of the form
\[
 \mathcal{S}^{\textup{nh}}(n;s,u)(t,b^1,\dots,b^m,w^1,\dots,w^{n-1})
 =
 X^{\textup{nh}}_{h_0}(s,u)t
 + \summ X^{\textup{nh}}_{h_i}(s,u)b^i
 + \summ X^{\textup{nh}}_{f_a}(s,u)w^a.
\]
We think of solutions of $\delta\Gamma =
\mathcal{S}^{\textup{nh}}(X,\Gamma)\delta X$ as 
non-holonomic diffusions.
This is in analogy to \cite[Chapter~V]{Bis81}
where Hamiltonian diffusions are considered in a similar manner.

For a function $f\in\cinf(K)$, viewed as a function on $TK$ via pull-back,
we have
\begin{align}\label{e:X_h_i}
 X^{\textup{nh}}_{h_0}(f)
 =
 -\by{1}{2}df.\mu_0^{-1}\I\summ\nabla^{\mathbb{I}}_{v_i}v_i,\text{ }
 X^{\textup{nh}}_{h_i}(f)
 =
 df.\mu_0^{-1}\I v_i,
 \textup{ and }
 X^{\textup{nh}}_{f_a}(f)
 =
 -df.\mu_0^{-1}\zeta_a.
\end{align}

Let $\Gamma$ be the solution semi-martingale to $\delta\Gamma =
\mathcal{S}^{\textup{nh}}(X,\Gamma)\delta X$ and let $\tau: TK\to K$
be the projection. 
Then $\tau\circ\Gamma$ solves
\[
 \delta(\tau\circ\Gamma)
 =
 T\tau.\delta\Gamma
 =
 -
 \by{1}{2}\mu_0^{-1}\I\summ\nabla^{\mathbb{I}}_{v_i}v_i(\tau\circ\Gamma)\delta t
 + \summ\mu_0^{-1}\I v_i(\tau\circ\Gamma)\delta B^i
 - \summ\mu_0^{-1}\zeta_a(\tau\circ\Gamma)\delta W^a.
\]
According to \cite[Chapter~V, Theorem~1.2]{IW89} this means that
the semi-martingale $\tau\circ\Gamma$ defines a diffusion in $K$ whose generator is 
the second order differential operator
\begin{equation}\label{e:A}
 -\by{1}{2}\mu_0^{-1}\I\summ\nabla^{\mathbb{I}}_{v_i}v_i
 + \by{1}{2}\summ(\mu_0^{-1}\I v_i)(\mu_0^{-1}\I v_i)
 + \by{1}{2}\summ(\mu_0^{-1}\zeta_a)(\mu_0^{-1}\zeta_a).
\end{equation}

To identify the drift of the diffusion $\tau\circ\Gamma$ a connection
is needed.
We
introduce the non-holonomic connection which is explained in
\cite[Section~5.1.1]{M02}:\footnote{Contrary to \cite{M02} we only use
  the projected version of the non-holonomic connection.}
Let $\textup{Pr}^{\mu}: TQ\to\D$ denote the projection onto $\D$ along
the $\mu$-orthogonal $\D^{\mu}$ of $\D$. Note that $\D^{\mu}\neq\ker
T\pi$ where $\pi: Q\toto K$ is the projection. Let 
$
 \textup{hl}^{\mathcal{A}}: \X(K)\to\X(Q;\D)
$, 
$\textup{hl}^{\mathcal{A}}(s,x)(u) = (s,u,x,-\A_s(u))$
be the horizontal lift map associated to $\A$. 
Given $X,Y\in\X(K)$ the non-holonomic connection is prescribed by
\begin{equation}\label{e:nh-conn}
 \nabla^{\textup{nh}}_XY
 =
 T\pi\textup{Pr}^{\mu}\nabla^{\mu}_{\textup{hl}^{\mathcal{A}}X}(\textup{hl}^{\mathcal{A}}Y).
\end{equation}
This connection is
metric, i.e., $\nabla^{\textup{nh}}\mu_0=0$, and  its geodesic equations
are exactly the equations of motion of the non-holonomic system
described by $(TK,\Omnh,\Hamc)$. However, in general,
$\nabla^{\textup{nh}}$ will have non-trivial torsion. 

\begin{lemma}
For $u,v\in\ko=\X_L(K)$ we have
\[
 \nabla^{\textup{nh}}_uv
 =
 \mu_0^{-1}(\I\nabla^{\mathbb{I}}_uv + \A^*\A[u,v])
\]
where 
$\nabla^{\mathbb{I}}_uv 
 =
 \by{1}{2}[u,v] + \by{1}{2}\I^{-1}([u,\I v]+[v,\I u])$.
Its torsion is given by 
$\textup{Tor}^{\textup{nh}}(u,v) = \mu_0^{-1}\A^*\A[u,v]$.
\end{lemma}

Note that $\nabla^{\textup{nh}}_uv$ is not left invariant any more.
At the compressed level the equations of motion of the Chaplygin ball
write as $u' + \mu_0^{-1}[u,\I u] = 0$. In 3D 
this corresponds to \cite[Equation~(3.5)]{D04}.

\begin{proof}
For $u,v\in\ko$ we need to compute
\[
 \textup{Pr}^{\mu}\nabla^{\mu}_{(u,-\A u)}(v,-\A v)
 =
 \textup{Pr}^{\mu}(\nabla^{\mathbb{I}}_uv - u.\A v)
 =
 \textup{Pr}^{\mu}(\nabla^{\mathbb{I}}_uv - \A[u,v]).
\]
Now note that $(w,X)\in\D^{\mu}$ if and only if
$w = \I^{-1}\A^* X$. We have to solve
\[
 0 
 =
 \A(\nabla^{\mathbb{I}}_uv + \I^{-1}\A^*X)
 - \A[u,v] + X
\]
for $\I^{-1}\A^*X$. The solution is found to be given by
$\mu_0\I^{-1}\A^*X = \A^*\A([u,v]-\nabla^{\mathbb{I}}_uv)$.
Therefore,
\begin{equation}
 \nabla^{\textup{nh}}_uv
 =
 \nabla^{\mathbb{I}}_uv 
 + \mu_0^{-1}\A^*\A([u,v]-\nabla^{\mathbb{I}}_uv)
 =
 \mu_0^{-1}\A^*\A[u,v]
 + \mu_0^{-1}\I\nabla^{\mathbb{I}}_uv.
\end{equation}
\end{proof}

Recall that the Hessian of $\nabla^{\textup{nh}}$ is defined by
\[
\textup{Hess}^{\textup{nh}}(f)(X,Y)
= 
XY(f)-\nabla^{\textup{nh}}_XY(f)
\]
for $X,Y\in\X(K)$ and  $f\in\cinf(K)$.

Let $f\in\cinf(K)$ and
$\delta\Gamma=\mathcal{S}^{\textup{nh}}(X,\Gamma)\delta X$.
By Proposition~\ref{prop:nh-Ito},
equations \eqref{e:X_h_i}
and 
$[dB^i,dB^j]=\delta^{i,j}dt$, 
$[dW^a,dW^b]=\delta^{a,b}dt$,
as well as
$[dB^i,dW^a]=0$
we have the Ito equation
\begin{align*}
 d(f\circ\Gamma)
 &=
 \big( 
  -\by{1}{2}\mu_0^{-1}\I\summ\nabla^{\mathbb{I}}_{v_i}v_i
     + \by{1}{2}\summ(\mu_0^{-1}\I v_i)(\mu_0^{-1}\I v_i)
     + \by{1}{2}\summ(\mu_0^{-1}\zeta_a)(\mu_0^{-1}\zeta_a)
 \big)f(\Gamma)dt\\
 &\phantom{++}
    + \summ(\mu_0^{-1}\I v_i)f(\Gamma)dB^i
    - \summ(\mu_0^{-1}\zeta_a)f(\Gamma)dW^a\\
 &=
 \big(
 -\by{1}{2}\mu_0^{-1}\I\summ\nabla^{\mathbb{I}}_{v_i}v_i
   + \by{1}{2}\summ\nabla^{\textup{nh}}_{\mu_0^{-1}\I v_i}(\mu_0^{-1}\I v_i)
   + \by{1}{2}\summ\nabla^{\textup{nh}}_{\mu_0^{-1}\zeta_a}(\mu_0^{-1}\zeta_a)
 \\
 &\phantom{++}
   + \by{1}{2}\summ\textup{Hess}^{\textup{nh}}(\mu_0^{-1}\I v_i,\mu_0^{-1}\I v_i)
   + \by{1}{2}\summ\textup{Hess}^{\textup{nh}}(\mu_0^{-1}\zeta_a,\mu_0^{-1}\zeta_a)
  \big) 
   f(\Gamma)dt\\
 &\phantom{++}
  + \summ(\mu_0^{-1}\I v_i)f(\Gamma)dB^i
  - \summ(\mu_0^{-1}\zeta_a)f(\Gamma)dW^a,
\end{align*}
which also confirms \eqref{e:A}.
Having split the generator into first and purely second order part it
makes sense to say what we mean by drift.

\begin{definition}\label{def:drift}
The vectorfield
\[
  -\by{1}{2}\mu_0^{-1}\I\summ\nabla^{\mathbb{I}}_{v_i}v_i
   + \by{1}{2}\summ\nabla^{\textup{nh}}_{\mu_0^{-1}\I v_i}(\mu_0^{-1}\I v_i)
   + \by{1}{2}\summ\nabla^{\textup{nh}}_{\mu_0^{-1}\zeta_a}(\mu_0^{-1}\zeta_a)
\]
is called the \caps{drift} of the diffusion $\tau\circ\Gamma$ with
respect to $\nabla^{\textup{nh}}$.
\end{definition}

Note that $\tau\circ\Gamma$ is a $\nabla^{\textup{nh}}$-martingale if
and only if the $\nabla^{\textup{nh}}$-drift vanishes. See also
\cite[Theorem~(7.31)]{E89}.

\begin{theorem}\label{thm:diff}
Let $\Gamma$ be a solution of the Stratonovich equation $\delta\Gamma =
\mathcal{S}^{\textup{nh}}(X,\Gamma)\delta X$ and let $\tau: TK\to K$
be the projection.
\begin{enumerate}[\up (1)]
\item
Then, with respect to the non-holonomic connection
$\nabla^{\textup{nh}}$ introduced in \eqref{e:nh-conn}, 
the semi-martingale $\tau\circ\Gamma$ defines a diffusion on
$K$ whose drift is the gradient $-\by{1}{2}\textup{grad}^{\mu_0}(\log\N)$ where $\N$ is
the density function defined in \eqref{e:N}.
\item
The drift $-\by{1}{2}\textup{grad}^{\mu_0}(\log\N)$ is horizontal with
respect to the mechanical connection,
$\textup{Hor}^{\textup{mech}} = (\ker T\kappa)^{\mu_0\bot}$,
on the principal
bundle $\kappa: K\toto K/H = S^{n-1}$.
If $\I$ satisfies the Hamiltonization condition \eqref{e:Ham-cond} then
the drift is also horizontal with respect to the
principal bundle connection 
$\summ\rho^{\alpha}\otimes Y_{\alpha}: TK\to\ho$
on $\kappa: K\toto K/H$. 
\end{enumerate}
\end{theorem}

Item (2) means that the drift's component of angular momentum about
the vertical axis in the space frame vanishes, and when the
Hamiltonization condition holds then the same is true for the
component of angular velocity about the vertical axis. For $n=3$ this
condition is always satisfied.


\begin{corollary}\label{cor:hom-case}
Let $\Gamma$ be a solution of the Stratonovich equation $\delta\Gamma =
\mathcal{S}^{\textup{nh}}(X,\Gamma)\delta X$ and let $\tau: TK\to K$
and
$\kappa: K\toto K/H = S^{n-1}$
be the obvious projections.
Suppose that  $\I = 1$ (i.e., the
ball is homogeneous).
\begin{enumerate}[\up (1)]
\item
Then $\tau\circ\Gamma$ defines a martingale in
$K$ with respect to the non-holonomic connection. 
\item
The process $\kappa\circ\tau\circ\Gamma$ is a Brownian motion on
$S^{n-1}$ whose generator is $\by{1}{2}$ times the Laplacian of $\nu$,
where $\nu$ is the metric on $S^{n-1}$ induced from the left
$H$-invariant metric $\mu_0 = \vv<(1+\A^*\A).,.>$ on $K$.
\end{enumerate} 
\end{corollary} 

Note that the restriction of $\mu_0$ to
$\hor=\hor^{\textup{mech}}=\textup{span}\set{\zeta_a}$ 
equals twice the restriction of the
biinvariant metric.
This corollary is intuitive but nevertheless not obvious since
the dynamics at the compressed level can never be described by a
Hamiltonian reduction procedure. This is because $\A$ is not the
mechanical connection, even if the ball is homogeneous. (Compare with
\cite[Corollary~4.3]{HG09}.) 
Thus it does
not fall in the category of \cite{LO08,LO08a}.
 

For reference we note the formula 
\begin{align}\label{e:v.mu_0}
 v.\mu_0(u)
 &=
 v.\summ\vv<\zeta_a,u>\zeta_a
 =
 \summ\vv<\zeta_a,[v,u]>\zeta_a
 - \summ\vv<\zeta_a,u>[v,\zeta_a]
 =
 \A^*\A[v,u] - [v,\A^*\A u].
\end{align}

\begin{proof}[Proof of Theorem~\ref{thm:diff}]
Let $f\in\cinf(K)$ which we regard via pull-back as a function on
$TK$. 
According to Definition~\ref{def:drift}
we need to show that 
\begin{equation}\label{e:0}
  -\mu_0^{-1}\I\summ\nabla^{\mathbb{I}}_{v_i}v_i
  + \summ\nabla^{\textup{nh}}_{\mu_0^{-1}\I v_i}(\mu_0^{-1}\I v_i)
  + \summ\nabla^{\textup{nh}}_{\mu_0^{-1}\zeta_a}(\mu_0^{-1}\zeta_a)
 =
 -\textup{grad}^{\mu_0}(\log\N).
\end{equation}
Claim:
\begin{equation}\label{e:1}
 \summ\nabla^{\textup{nh}}_{\mu_0^{-1}\I v_i}(\mu_0^{-1}\I v_i)
  =
 \mu_0^{-1}\I\summ\nabla^{\mathbb{I}}_{v_i}v_i
 -\textup{grad}^{\mu_0}(\log\N).
\end{equation}
Indeed,
we use \eqref{e:v.mu_0} and the fact that $u = \summ\vv<\I v_i,u>v_i$
to see that 
\begin{align*}
 \summ\nabla^{\textup{nh}}_{\mu_0^{-1}\I v_i}(\mu_0^{-1}\I v_i)
 &=
 \summ
  \vv<\I\mu_0^{-1}\I v_i,v_j>\vv<\I\mu_0^{-1}\I v_i,v_k>\nabla^{\textup{nh}}_{v_j}v_k
  + 
  \summ 
  \vv<\I\mu_0^{-1}\I v_i,v_j>\big(v_j.\vv<\I\mu_0^{-1}\I v_i,v_k>\big)v_k\\
 &=
 \summ\mu_0^{-1}[\mu_0^{-1}\I v_i,\I\mu_0^{-1}\I v_i]
 +
 \summ 
  \vv<\I\mu_0^{-1}\I v_i,v_j>\vv<\I\mu_0^{-1}[v_j,\A^*\A\mu_0^{-1}\I v_i],v_k>v_k\\
 &=
 \summ\mu_0^{-1}[\mu_0^{-1}\I v_i,\I\mu_0^{-1}\I v_i]
 +
 \summ\mu_0^{-1}[\mu_0^{-1}\I v_i,\A^*\A\mu_0^{-1}\I v_i]\\
 &=
 \summ\mu_0^{-1}[\mu_0^{-1}\I v_i,\I v_i]\\
 &=
 \summ\mu_0^{-1}[v_i,\I v_i]
  - 
  \summ\mu_0^{-1}[\mu_0^{-1}\A^*\A v_i,\I v_i] 
\end{align*}
where we use $\mu_0^{-1}\I = 1 - \mu_0^{-1}\A^*\A$ in the last equation.
Notice that $[v_i,\I v_i] = \I\nabla^{\mathbb{I}}_{v_i}v_i$.
For the gradient part of claim \eqref{e:1} we consider
\begin{align*}
 \vv<\summ[\mu_0^{-1}\A^*\A v_i,\I v_i],\zeta_b>
 &=
 -\summ\vv<\I v_i,[\mu_0^{-1}\vv<\zeta_a,v_i>\zeta_a,\zeta_b]>
 =
 -\summ\vv<\zeta_a,\vv<\I v_i,[\mu_0^{-1}\zeta_a,\zeta_b]>v_i>\\
 &=
 \summ\vv<[\mu_0^{-1}\zeta_a,\zeta_a],\zeta_b>
 =
 d(\log\N)\zeta_b 
\end{align*}
by \eqref{e:dN}.
Similarly it is true that 
$\vv<\summ[\mu_0^{-1}\A^*\A v_i,\I v_i],\xi_{\alpha}>
 =  \summ\vv<[\mu_0^{-1}\zeta_a,\zeta_a],\xi_{\alpha}>$. 
Using the property 
$[Z_a,Y_{\alpha}] = \delta_{a,b}Z_c - \delta_{a,c}Z_b$ with
$Y_{\alpha} = [Z_b,Z_c]$ 
it is easy to see that 
\[ 
 \summ\vv<[\mu_0^{-1}\zeta_a,\zeta_a],\xi_{\alpha}>
 = 
 \vv<\mu_0^{-1}\zeta_b,\zeta_c>
 -
 \vv<\mu_0^{-1}\zeta_c,\zeta_b>
 =
 0.
\]
Thus
$\summ\mu_0^{-1}[\mu_0^{-1}\A^*\A v_i,\I v_i] 
 = \mu_0^{-1}\summ (d(\log\N)\zeta_b)\zeta_b
 = \textup{grad}^{\mu_0}(\log\N)$
and claim \eqref{e:1} follows.

Claim:
\begin{equation}\label{e:2}
 \nabla^{\textup{nh}}_{\mu_0^{-1}\zeta_a}(\mu_0^{-1}\zeta_a) = 0.
\end{equation}
Again we use \eqref{e:v.mu_0} to see that
\begin{align*}
 \nabla^{\textup{nh}}_{\mu_0^{-1}\zeta_a}(\mu_0^{-1}\zeta_a)
 &=
 \mu_0^{-1}[\mu_0^{-1}\zeta_a,\I \mu_0^{-1}\zeta_a] 
 + 
 \summ
  \vv<\I v_j,\mu_0^{-1}\zeta_a>\big(v_j.\vv<\I v_k,\mu_0^{-1}\zeta_a>\big)v_k\\
 &=
 \mu_0^{-1}[\mu_0^{-1}\zeta_a,\I \mu_0^{-1}\zeta_a] 
 +
 \summ
  \vv<\I v_j,\mu_0^{-1}\zeta_a>\vv<\I v_k,\mu_0^{-1}[v_j,\A^*\A\mu_0^{-1}\zeta_a]>v_k\\
 &\phantom{++}
 -
 \summ
  \vv<\I v_j,\mu_0^{-1}\zeta_a>\vv<\I v_k,\mu_0^{-1}[v_j,\zeta_a]>v_k\\
 &=
 \mu_0^{-1}[\mu_0^{-1}\zeta_a,\I \mu_0^{-1}\zeta_a] 
 +
 \mu_0^{-1}[\mu_0^{-1}\zeta_a,\A^*\A\mu_0^{-1}\zeta_a] 
 -
 \mu_0^{-1}[\mu_0^{-1}\zeta_a,\zeta_a]\\
 &= 
 0. 
\end{align*}    
Now \eqref{e:1} and \eqref{e:2} imply \eqref{e:0} which shows part (1)
of the assertion.

For (2) one checks that \eqref{e:Ham-cond} yields, for 
$\xi_{\alpha} = [\zeta_c,\zeta_d]$, 
\begin{align*}
 \vv<\textup{grad}^{\mu_0}(\log\N),\xi_{\alpha}> 
 &=
 \summ
 \vv<\mu_0^{-1}[\zeta_c,\zeta_d],\vv<[\mu_0^{-1}\zeta_a,\zeta_a],\zeta_b>\zeta_b>\\
 &=
 \summ
 \vv<\mu_0^{-1}[\zeta_c,\zeta_d],\zeta_b>
  \vv<\mu_0^{-1}\zeta_a,[\zeta_a,\zeta_b]>\\
 &=
 \vv<\mu_0^{-1}\zeta_d,[\zeta_c,\zeta_d]>\vv<\mu_0^{-1}\zeta_c,[\zeta_c,\zeta_d]>(n-2)
 +
 \vv<\mu_0^{-1}\zeta_c,[\zeta_c,\zeta_d]>\vv<\mu_0^{-1}\zeta_d,[\zeta_d,\zeta_c]>(n-2)\\
 &=
 0
\end{align*}  
where we also have used that 
$\summ\vv<[\mu_0^{-1}\zeta_a,\zeta_a],\xi_{\alpha}> = 0$.
\end{proof}

\begin{proof}[Proof of Corollary~\ref{cor:hom-case}]
Part (1) is clear.

Concerning part (2) let $f\in\cinf(S^{n-1})$. According to
Definition~\ref{def:BM} we should to show that the generator
\eqref{e:A} satisfies
\[
 \big(
 \by{1}{2}\summ(\mu_0^{-1}v_i)(\mu_0^{-1}v_i)
 + \by{1}{2}\summ(\mu_0^{-1}\zeta_a)(\mu_0^{-1}\zeta_a)
 \big)\kappa^*f
 =
 \by{1}{2}\kappa^*\Delta^{\nu}f
\]
where $\Delta^{\nu}$ is the Laplacian associated to $\nu$. Indeed, we find
$ 
\by{1}{2}\summ(\mu_0^{-1}v_i)(\mu_0^{-1}v_i)\kappa^*f
 + \by{1}{2}\summ(\mu_0^{-1}\zeta_a)(\mu_0^{-1}\zeta_a)\kappa^*f
=
\by{2}{2}\by{1}{4}\summ\zeta_a\zeta_a\kappa^*f$.
Now $\set{\by{1}{\sqrt{2}}\zeta_a}$ is a horizontal orthonormal frame for
$\mu_0|(\hor\times\hor)$ where $\hor$ is the $\mu_0$-orthogonal to $\ker
T\kappa$. 
Therefore,
\[
\by{1}{4}\summ\zeta_a\zeta_a\kappa^*f
= \by{1}{2}\textup{Tr}^{\textup{hor}}\textup{Hess}^{\mu_0}(\kappa^*f)
= \by{1}{2}\kappa^*\Delta^{\nu}f
\]
where $\textup{Tr}^{\textup{hor}}$ denotes the trace computed with
respect to horizontal fields only
and $\textup{Hess}^{\mu_0}$ is the Hessian of the Levi-Civita
connection on $(K,\mu_0)$. 
The equation 
$
\zeta_a\zeta_a\kappa^*f
= \textup{Hess}^{\mu_0}(\kappa^*f)(\zeta_a,\zeta_a)
$
is justified by the observation that 
$\nabla^{\mu_0}_{\zeta_a}\zeta_a = 0$
where $\nabla^{\mu_0}$ is the  
Levi-Civita connection of the right
invariant metric $\mu_0$.
(In fact, the restriction of $\nabla^{\textup{nh}}$ to
$\hor\times\hor$ equals the restriction of $\nabla^{\mu_0}$.)
\end{proof}

In the homogeneous case the above construction yields a
Brownian motion on $S^{n-1}$ in a manner similar to the one described in
\cite[Chapter~V]{IW89} by the notion of rolling the sphere $S^{n-1}$ along a
Brownian motion in $\R^{n-1}$ by means of the Levi-Civita connection. 
The difference is that \cite{IW89} start from Brownian motion in
$\R^{n-1}$ while we started from Brownian motion in
$\R^m\times\R^{n-1}$ with $m=\dim\so(n)=\by{n(n-1)}{2}$.
One can recover the setting of \cite{IW89} by setting
$(H_0,H_1,\dots,H_m)=0$ in \eqref{e:h-f}. Then, with $\I=1$, we obtain
a diffusion $\kappa\circ\tau\circ\Gamma$ 
which
is driven by Brownian motion $(W^1,\dots,W^{n-1})$ in $\R^{n-1}$ and  the
generator of which is given by
\[
 \by{1}{2}\summ(\mu_0^{-1}\zeta_a)(\mu_0^{-1}\zeta_a)\kappa^*f
 =
 \by{1}{2}\kappa^*\Delta^{\nu}f
\]
where $f\in\cinf(S^{n-1})$. 
Referring to the interpretation stated in the introduction this means
that the Chaplygin ball is subjected to horizontal jiggling but there
is no angular jiggling. Equivalently, the ball sits on a table which
undergoes a translational Brownian motion. 
Compare \cite{MS90}.
Alternatively, we can set $(F_1,\dots,F_{n-1})=0$ in
\eqref{e:h-f}. Then there is only angular jiggling and the diffusion
$\tau\circ\Gamma$ is driven by $(B^1,\dots,B^m)$. By \eqref{e:2} the
drift remains the same as in Theorem~\ref{thm:diff}.


It seems that the notion of a stochastic non-holonomic system has been
hardly investigated in the literature. 
We finish by asking the following questions.

\begin{enumerate}[\up (1)]
\item
Does an analog of Theorem~\ref{thm:diff} hold for general
$G$-Chaplygin systems when there is a preserved measure? What can be
said about the drift if there is no preserved measure?
\item
Which is the precise relationship between $G$-Chaplygin systems with
preserved measures and measure preserving `Chaplygin diffusions'?
Preservation of measure by diffusions is studied in \cite[Chapter~V]{IW89}.
\item
Is there a time change or Girsanov type argument to eliminate the drift in
Theorem~\ref{thm:diff} or to make it even a Brownian motion? 
Is this related to the Hamiltonization of the
deterministic problem? 
\end{enumerate}

\section{Appendix: $G$-Chaplygin systems and symmetry reduction}\label{sec:Chap}

The purpose of this appendix is to shortly introduce and motivate the
notion of a $G$-Chaplygin system and to state
Proposition~\ref{prop:comp} which explains the symmetry reduction of
such systems.
This reduction is termed \emph{compression}
(\cite{EKMR04}) to distinguish it from
symplectic reduction. These concepts are closely related and
compression can be viewed as a perturbed version of its symplectic
counterpart. At the same time, however, there are fundamental
differences; symmetries behave differently in non-holonomic mechanics
and do not necessarily give rise to conserved quantities, and there need
not exist a preserved measure (\cite[Section~5.4]{M02}); all this is
related to the question of closedness of the form $\Omnh$ defined in
Proposition~\ref{prop:comp}. See \cite{BS93,B03,M02,EKMR04,HG09}.

A non-holonomic system is a triple $(Q,\D,L)$
where $Q$ is a configuration manifold,
$L: TQ\to\R$
is  a Lagrangian, and  $\D\subset TQ$ is
a smooth non-integrable distribution which is supposed to be
of constant rank. The equations of motion for a
curve $q(t)$ which should satisfy $q'\in\D$ are then stated
in terms of the Lagrange d'Alembert principle. 
Suppose there is a Riemannian metric $\mu$ on $Q$ such that we have an
isomorphism $TQ\cong T^*Q$ and assume that $L$ is the kinetic energy Lagrangian.
In this
case there is also an (almost) Hamiltonian version: 
continue to use the symbol $\mu$ to
denote the co-metric and consider the Hamiltonian $\Ham(q,p)$
given by the Legendre transform of $L$.  Since $\D$ is of constant rank
there is a family of
independent one-forms $\phi^a\in\Om(Q)$ such that $\D$ is
the joint kernel of these.  In terms of coordinates
$(q^i,p_i)$ the equations of motion are
\[
 (q^i)' = \dd{p_i}{\mathcal{H}}
 \textup{ and }
 p_i' =
  -\dd{q^i}{\mathcal{H}} - \summ\lam_a\phi^a(\dd{q^i}{})
\]
where the $\lam_a$ are the Lagrange multipliers to be determined from
the supplementary condition that  $\mu(\phi^a,p)=0$.
With $X^{\mathcal{C}}_{\mathcal{H}} := (q',p')$ we may
thus rephrase the equations as
\begin{equation}\label{e:eom-with-lam}
 i(X^{\mathcal{C}}_{\mathcal{H}})\Om = d\Ham + \summ\lam_a\tau^*\phi^a
\end{equation}
where $\Om=-d\theta$ is the canonical symplectic form on
$T^*Q$ and $\tau: T^*Q\to Q$ is the footpoint projection.
(The notation $X^{\mathcal{C}}_{\mathcal{H}}$ will become clear below.)


Let $G$ be a Lie group that
acts freely, properly and by isometries on the Riemannian
manifold $(Q,\mu)$. A $G$-\caps{Chaplygin system} is a
non-holonomic system $(Q,L=\by{1}{2}||\cdot ||^{2}_{\mu},\D)$ that
has the property that $\D$ is a
principal connection on the principal bundle $Q\toto
Q/G$. Thus $\D$ is the kernel of a connection form $\A:
TQ\to\gu$. Notice that we do not require $\A$ to be the
mechanical connection associated to $\mu$. 

Consider $\C$ and $\Om^{\mathcal{C}}$ as defined in
Section~\ref{sec:stoch-Chap}. 
Since $X^{\mathcal{C}}_{\mathcal{H}}$ is, by construction, tangent to $\M$
and takes values in $\C$ one may now rewrite the equations
of motion \eqref{e:eom-with-lam} in the appealing format
\[
 i(X^{\mathcal{C}}_{\mathcal{H}})\Om^{\mathcal{C}}
 =
 (d\Ham)^{\mathcal{C}}
\]
where $(d\Ham)^{\mathcal{C}}$ is the restriction of
$\iota^*d\Ham$ to $\C$ with $\iota: \D\hookto TQ$ being the
inclusion.


Let $\mu_0$ denote the induced metric on $S:=Q/G$ that
makes $\pi: Q\toto S$ a Riemannian submersion. 
Identify tangent and
cotangent space of $Q$ and $S$ via their respective
metrics.
Consider the orbit projection map
\[
 \rho := T\pi|\D: \D\toto \D/G = TS.
\]
We may also associate a fiber-wise inverse to this
mapping which is given by the horizontal lift mapping
$\hlphi$ associated to $\A$. 
As already noted in Section~\ref{sec:stoch-Chap}, 
$\iota^*\tau^*\A: T\D\to\gu$ 
defines a principal bundle connection
for $\rho$, whose horizontal space is given by $\C$.

\begin{proposition}[Compression]\label{prop:comp}
The following are true.
\begin{enumerate}[\up (1)]
\item
$\Om^{\mathcal{C}}$ descends to a
non-degenerate two-form $\Omnh$ on $TS$.
\item
$\Omnh = \Om_S - \vv<J_G\circ\textup{hl}^{\A},\tau_S^*\curv^{\mathcal{A}}>$.
Here
$\Om_S=-d\theta_S$ is the canonical form on $TS$, $J_G$ is
the \momap of the tangent lifted $G$-action on $TQ$,
$\curv^{\mathcal{A}}\in\Om^2(S,\gu)$ is the curvature form of $\A$, and
$\tau_S: TS\to S$ is the projection.
\item
Let $h: TQ\to\R$ be $G$-invariant. Then
the vectorfield
\[
X^{\mathcal{C}}_h:=(\Om^{\mathcal{C}})^{-1}(dh)^{\mathcal{C}},
\]
where $(dh)^{\mathcal{C}}$ is the restriction of
$\iota^*dh$ to $\C$,
is $\rho$-related
to the vector field $X^{\textup{nh}}_{h_0}$ on $T^*S$ defined by
\[
 i(X^{\textup{nh}}_{h_0})\Omnh = dh_0
\] 
where the compressed
Hamiltonian, $h_0: T^*S = TS \to\R$ is defined by $h_0
:= h\circ\iota\circ\textup{hl}^{\A}$, with $\textup{hl}^{\A}$
denoting the horizontal lift mapping.
\end{enumerate}
\end{proposition}

In general, $\Omnh$ is an almost symplectic form, that is, it is
non-degenerate and non-closed. See \cite{BS93,EKMR04,HG09}.

\end{document}